\documentclass[aps,pra,reprint,showpacs,superscriptaddress,twocolum]{revtex4-1}
\usepackage{amsmath,amssymb,graphicx}

\newcommand{\bitem}{\begin{itemize}}
\newcommand{\fitem}{\end{itemize}}
\newcommand{\beq}{\begin{equation}}
\newcommand{\eeq}{\end{equation}}
\newcommand{\beqa}{\begin{eqnarray}}
\newcommand{\eeqa}{\end{eqnarray}}

\begin{document}
	
\title{Dissipative phonon Fock state production in strong nonlinear optomechanics}

\author{G. D. de Moraes Neto} \affiliation{Institute of Fundamental and Frontier Sciences, University of Electronic Science and Technology of China, Chengdu, PR China}\affiliation{Instituto de F\'{i}sica, Universidade Federal de Uberl\^{a}ndia, Uberl\^{a}ndia, Minas Gerais 38400-902, Brazil}

\author{{V. Montenegro}} \affiliation{Institute of Fundamental and Frontier Sciences, University of Electronic Science and Technology of China, Chengdu, PR China}

\author{V. F. Teizen} \affiliation{Instituto de F\'{i}sica de S\~ao Carlos, Universidade de S\~ao Paulo, 13560-970 S\~ao Carlos, SP, Brazil}

\author{E. Vernek} \affiliation{Instituto de F\'{i}sica, Universidade Federal de Uberl\^{a}ndia, Uberl\^{a}ndia, Minas Gerais 38400-902, Brazil}

\begin{abstract}
	We put forward a deterministic dissipative protocol to prepare phonon Fock states in nonlinear quantum optomechanical devices. The system is composed of a mechanical mode interacting with an optical field via radiation pressure, whereas the light mode is laser-driven in the resolved blue-sideband regime. To keep our results tractable, we have switched to an interaction picture in a displaced basis, where the effective Hamiltonian exhibits the selective photon-phonon interaction explicitly. After proper parameter adjustment and similarly to cavity-cooling schemes, the quantum evolution allows steering the mechanical degree of freedom to the desired Fock state by directing the optical excitations dynamically towards the target phonon state. The numerical results, including decoherence on both the mechanical and the optical degrees of freedom, show to be quite robust in the good- and bad-cavity regimes with fidelities exceeding 95\%. Lastly, characterization of the achieved nonclassicality, as well as the limitations and feasibility of our protocol under experimental parameters, are also analyzed. 
	%
\end{abstract}

	\maketitle
	\section{Introduction}
	The thorough investigation of nonclassical states has proven to be of the utmost importance for fundamental and experimental applications on related quantum topics \cite{RomeroIsart2011, Bose2017}. For instance, to examine the interface between the quantum-to-classical transitions \cite{ChengHua2017}, or to provide a useful resource in the tireless quest of a theory for quantum gravity \cite{Bose2017}. Furthermore, in the quantum information arena, the successful advent of quantum computation and quantum communication fields entail long-lived quantum states as well as quantum correlations \cite{Livro, fiber}, crucial to surpass its classical counterparts \cite{Preskill2012, Boixo2018}. Nonetheless, the inescapable sources of noise and decoherence in the quantum evolution make the production of long-lived quantum states considerably challenging \cite{Zugenmaier2018experimental}. Nowadays, substantial efforts have been devoted to promote efficient techniques for preparing and protecting nonclassical states from quantum noise \cite{Suter2016, Duan1997}, for instance, decoherence-free subspaces \cite{DFS, destorsz}, dynamical decoupling \cite{DD}, and reservoir engineering \cite{PCZ, RE, PNS}, etc.
	
	In this context, the field of quantum optomechanics \cite{Optomechrev} emerges as a formidable platform to accomplish, for example, generation of quantum states for the light and/or the matter degrees of freedom \cite{Bose1997preparation, Optomechrev, trampoptmech, Verhagen}. The ready access to nonlinear (trilinear) single-photon interaction between micro- and nano-fabricated mechanical resonators and the optical degrees of freedom makes the production of quantum states experimentally available in the weak to strong optomechanical regimes \cite{Optomechrev}. Moreover, current schemes to realize quantum state tomography of mechanical resonators \cite{Vanner2011}, and the ability to cool the phononic excitations down to its ground state \cite{Vanner2013, OConnell2010, Ojanen2014, Martin2014, Rao2016} provide a fertile ground to produce phonon states in a controllable fashion. In particular, the predominant schemes to prepare single phonon excitations \cite{Fock1}, squeezed \cite{Squee} and Schr\"{o}dinger cat states \cite{Bose1997preparation} are intrinsically probabilistic, as they are mainly based on measuring the optical mode (correlated with the vibrational mode), thus collapsing the vibrational modes into a nonclassical states \cite{Latmiral2018}. Nevertheless, deterministic schemes can also be achieved by steering the system towards a stationary state, the so-called reservoir engineering protocol \cite{PCZ}. As stated earlier, such protocol not only serves as a mechanism to bypass quantum decoherence, but also it is potentially useful to prepare superpositions of two wave packets \cite{SP,SP1}. This technique \cite{PCZ}, experimentally demonstrated in a trapped ion system \cite{IT}, signals a step of paramount importance towards the implementation of quantum information resources. The proposal can accomplish goals such as dissipative preparation of many-body quantum states \cite{Many}, universal dissipative quantum computation \cite{UniDissQC}, and analog quantum simulation in open systems \cite{ExpDissEng} allowing studies on quantum phase transitions. Naturally, one key aspect of dissipative protocols is their independence on initial states. In other words, it is possible to construct, from an arbitrary initial state, a non-unitary dynamic which can generate a steady state that asymptotically approaches to some desired target state. 
	
	This work is devoted to the deterministic generation of phononic Fock states in laser-driven nonlinear quantum optomechanics. The system, operating in the single-photon optomechanical strong regime \cite{chinesreview}, is depicted in a standard Fabry-Perot configuration in Fig. \ref{fig:fig1}. After justified approximations and switching into an interaction picture within a displaced mechanical basis, we succeed into deriving an effective Hamiltonian, where its sole representation allows us to unthread the physical mechanisms. We show that, once the parameters are accurately tuned, it is possible to generate a steady-state of a single $|M\rangle$ phonon Fock state, by transferring the photonic excitations towards the targeted phonon state. In this manner, our dissipative scheme is related to the optomechanical cavity-cooling protocol ---a setup which has brought recently mechanical resonators to their ground states--- driven by a nonlinear quantum scissor \cite{scissors, Rafael}. We present our findings both in the bad-cavity regime, for which we have obtained an effective master equation in Lindblad form, as well as for the good-cavity regime (no closed analytic form was found). Furthermore, when including sources of decoherence, our scheme shows to be quite robust, with production fidelities exceeding $95 \%$. We stress that the nonlinear optomechanical coupling strength is the main parameter for a plethora of proposals, for instance, to generate nonclassical states of photons and phonons \cite{NooN,PRA.2013.88.063819,PRA.2013.87.053849}, as well as to observe photon blockade effect \cite{block}. 
	
	In the next section we will derive an effective evolution for the strong nonlinear optomechanical coupling regime in the bad-cavity limit, for which no-linearization of the optomechanical system is performed. In the third section, we will show that our protocol works in the bad cavity regime and numerically confirm that remains valid even when operating in the good-cavity regime. Finally, section 4 is dedicated to our concluding remarks.

	\section{optomechanical dynamics}
	\begin{figure}
		\centering \includegraphics[width=0.75\linewidth]{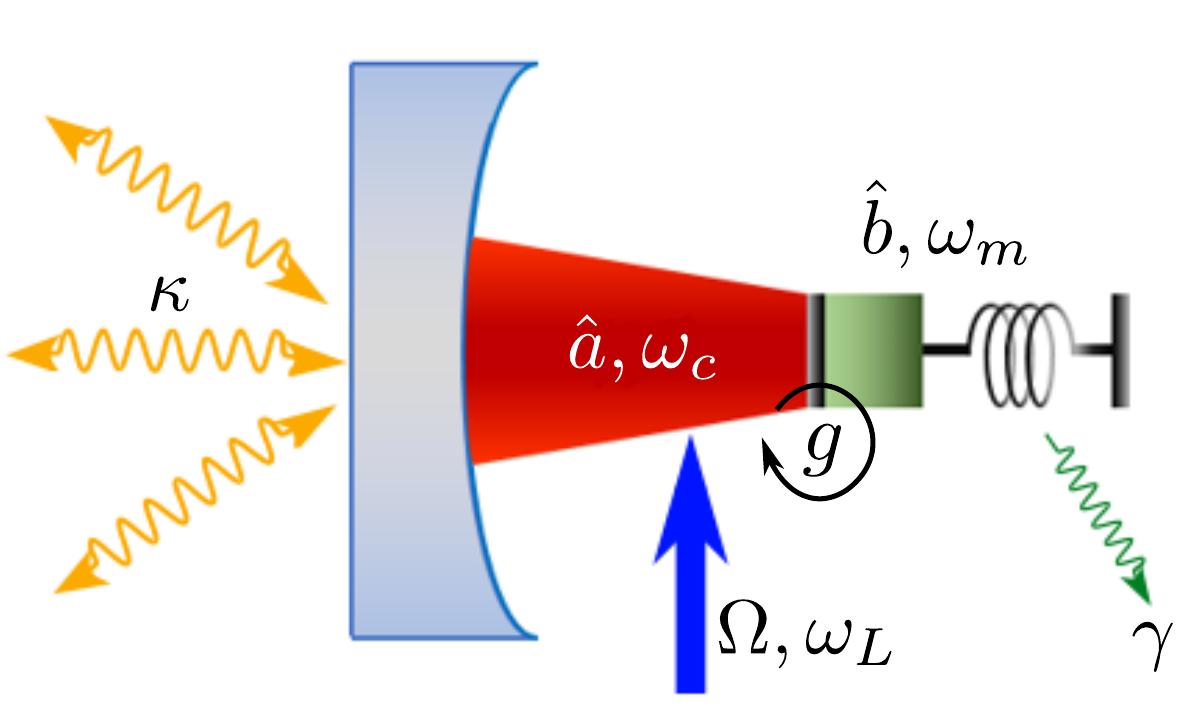}
		\caption{Sketch of an open optomechanical system driven by an external laser $(\Omega, \omega_L$). The optical mode ($\hat{a}, \omega_c$) is nonlinearly coupled ($g$) to a mechanical oscillator ($\hat{b}, \omega_m$); $\kappa$ ($\gamma$) stands for the cavity (mechanical) decay (damping) rate.}
		\label{fig:fig1}
	\end{figure}
	
	For the sake of clarity, we will briefly include the derivation of the main Hamiltonian in this section. These steps can be followed in more detail with the aid of Refs. \cite{Velazquez2015, Bose1997preparation}, for example.

	We study a standard driving optomechanical system composed of a mechanical mode of frequency $\omega_{m}$ coupled to a cavity mode of frequency $\omega_{c}$ via radiation-pressure interaction. Additionally, an external laser drives the optical mode with angular frequency $\omega_L$ and laser amplitude $\Omega = \Omega^*$, as schematically depicted in Fig. \ref{fig:fig1}. Hence, the optomechanical Hamiltonian is ($\hbar =1$)
	
	\begin{equation}
	\hat{H} = \omega _{c}\hat{a}^{\dagger }\hat{a}+\omega _{m}\hat{b}^{\dagger} \hat{b}-g\hat{a}^{\dagger }\hat{a}(\hat{b}^{\dagger} + \hat{b}) + [\Omega e^{-i \omega _{L}t}\hat{a}^{\dagger } + \mathrm{H.c}],\label{cell}
	\end{equation}
	where, $\hat{b}$ ($\hat{a}$) is the usual annihilation boson operator for the mechanical (optical) mode and $g$ represents the single-photon coupling strength.
	
	Firstly, let us eliminate the time dependence from Eq. (\ref{cell}) by moving to a rotating frame at the external laser frequency, transforming as
	
	\begin{equation}
	\hat{H} = -\Delta\hat{a}^{\dagger}\hat{a}+\omega_{m}\hat{b}^{\dagger} \hat{b}-g\hat{a}^{\dagger }\hat{a}(\hat{b}+\hat{b}^{\dagger}) + \Omega(\hat{a} + \hat{a}^\dagger),
	\end{equation}
	where $\Delta = \omega_L - \omega_c$ is the pump detuning relative to the cavity frequency. Secondly, we can obtain physical insights by moving to the optomechanical displaced basis that diagonalizes the radiation pressure interaction. This can be achieved with the help of the displacement operator $\hat{D}(\hat{\xi}) = e^{\hat{\xi} \hat{b}^{\dagger}-\hat{\xi}^{\dagger} \hat{b}}$, in which $\hat{\xi} = \hat{a}^{\dagger}\hat{a} g/\omega_m$. To obtain the modified Hamiltonian in the joint basis, we can evoke the Baker–Campbell–Hausdorff formula \cite{Bose1997preparation}, that allows us to write $\hat{D}^\dagger(\hat{\xi}) \hat{b} \hat{D}(\hat{\xi}) = \hat{b} + g/\omega_m \hat{a}^\dagger\hat{a}, \hat{D}^\dagger(\hat{\xi}) \hat{a} \hat{D}(\hat{\xi}) = \hat{a}\hat{D}(g/\omega_m),$ and $\hat{D}^\dagger(\hat{\xi}) \hat{a}^\dagger\hat{a} \hat{D}(\hat{\xi}) = \hat{a}^\dagger\hat{a}.$ Furthermore, by considering an interaction picture with unitary transformation $\hat{U} = \mathrm{exp}[-i(-\Delta\hat{a}^{\dagger }\hat{a}+\omega _{m}\hat{b}^{\dagger }\hat{b})t]$ and using the similarity transformation, i.e., the fact that for any function $f$, unitary operator $\hat{u}$, with arbitrary set operators $\{\hat{X}_i\}$ it holds that $\hat{u} f(\{\hat{X}_i\}) \hat{u}^\dagger = f(\{\hat{u}\hat{X}_i\hat{u}^\dagger\})$ (see Appendix in Ref. \cite{Bose1997preparation}), we can readily write 
	
	\begin{equation}
	\hat{H}_{D} = -\frac{g^{2}}{\omega _{m}}(\hat{a}^{\dagger }\hat{a} )^{2}+ \left[\Omega e^{-i\Delta t}\hat{a}^{\dagger} \hat{D}^\dagger(\eta e^{i\omega _{m}t}) + \mathrm{H.c}\right], \label{eq:Hd}
	\end{equation}
	where $\eta = g/\omega_m$ is the scaled optomechanical coupling interaction. As pointed out in Ref. \cite{Velazquez2015}, the second term of Eq. (\ref{eq:Hd}) resembles that of a driven trapped ion. Thus, it suggests to follow a similar approach as the one used in trapped-ion QED. Hence, we proceed to expand the mechanical displacement operator $\hat{D}(\hat{\xi})$ into their power series. $\hat{D}(\eta e^{i\omega_m t}) = e^{-\eta^2/2}\sum_{p,q=0}^\infty 1/(p!q!)(\eta \hat{b}^\dagger)^p (-\eta \hat{b})^q e^{-i\omega_m(q-p)t}.$ Switching to an adequate interaction picture, such as we can eliminate the quadratic Kerr-like term in Eq. (\ref{eq:Hd}), and using the commutation relation $f(\hat{n})\hat{a}^\dagger = \hat{a}^\dagger f(\hat{n} + 1)$, we can obtain $e^{-i g\eta t \hat{n}^2} \hat{a}^\dagger e^{i g\eta t \hat{n}^2} = e^{-i g\eta t (2\hat{n} + 1)} \hat{a}^\dagger$. With this, the Hamiltonian in Eq. (\ref{eq:Hd}) acquires the form
	
	\begin{equation}
	\hat{H}_D = \Omega e^{-i\Delta t} e^{-i g\eta t (2\hat{n} + 1)} \hat{a}^\dagger\hat{D}^\dagger(\eta e^{i\omega_m t})  + \mathrm{H.c.} \label{eq:noapprox}
	\end{equation}
	
	Note that, to obtain Eq. (\ref{eq:noapprox}), no approximation has been made so far. In the following, we will consider the resolved sideband regime $\kappa \ll \omega_m$, an operational regime typically used in current optomechanical protocols, where the cavity bandwidth is small when compared to the mechanical resonance frequency. This regime guarantees that the optical cavity may be employed as a frequency-selective element for performing coherent control, but it also limits the quantity of circulating optical power. Specifically, we consider the blue-sideband in the single-photon subspace, i.e., $\Delta + g\eta = s\omega_m$, being $s = \{0,1,2\ldots \}$. Moreover, we will also consider laser intensity sufficiently low $\Omega \ll \omega_m$. Finally, by invoking the rotating-wave approximation, we can neglect higher frequencies in the quantum dynamics and obtain
	
	\begin{eqnarray}
	\hat{H}_\mathrm{RWA} &=& \Omega e^{-\frac{\eta^2}{2}}|0\rangle\langle 1|\frac{(\hat{b}^\dagger\hat{b})!}{(\hat{b}^\dagger\hat{b} + s)!}L_{\hat{b}^\dagger\hat{b}}^{(s)}(\eta^2)(\eta \hat{b})^s + \mathrm{H.c.}
	\end{eqnarray}
	
	It is straightforward to write the Hamiltonian for the special case of $s = 1$
	
	\begin{equation}
	\hat{H}_\mathrm{eff} = |0\rangle\langle 1| \hat{\chi}^{(1)}(\eta)\hat{b} + |1\rangle\langle 0| \hat{b}^\dagger \hat{\chi}^{(1)}(\eta), \label{effective}
	\end{equation}
	where the operator $\hat{\chi}^{(1)}(\eta)$ is
	
	\begin{equation}
	\hat{\chi}^{(1)}(\eta) = \eta \Omega e^{-\frac{\eta^2}{2}} \frac{L_{\hat{b}^\dagger\hat{b}}^{(1)}(\eta^2)}{\hat{b}^\dagger\hat{b} + 1},
	\end{equation}
	and $L_{n}^{m}(x)$ are associated Laguerre polynomials.
	
	Expanding the operators $\hat{b}$ and $\hat{b}^{\dagger}$ in the Fock space basis and adjusting the parameter $\eta$, such as $L_{M}^{(1)}(\eta^{2}) = 0$, we can safely state that for any state $\left\vert photon, phonon\right\rangle$ we have $\hat{H}_\mathrm{eff} \left \vert N, M \right\rangle = 0$ and consequently, for an initial vibrational state prepared within the \emph{upper-bound} $\mathrm{(ub)}$ subspace ranging from $\left\vert 0 \right \rangle $ to $\left\vert M \right\rangle$, the Hamiltonian $\hat{H}_\mathrm{eff}$ becomes $\hat{H}_\mathrm{eff}^{\mathrm{(ub)}} = \hat{B} |0\rangle \langle 1| + \hat{B}^\dagger |1 \rangle\langle 0|,$ where $\hat{B} = \sum_{m=0}^{M-1}\hat{\chi}^{(1)}(\eta) \sqrt{m+1} \left \vert m \right\rangle \left\langle m + 1\right \vert$. At this stage, we can notice the importance of the derivation of the effective Hamiltonian in Eq. (\ref{effective}), as the suppression of the Laguerre polynomial gives us the precise optomechanical coupling $\eta$ for a given $M$. Furthermore, the explicit combination between a confined $|0\rangle$ and $|1\rangle$ photonic manifold (photon blockade effect) and the production of a phononic dark state within a sliced subspace $0 \leq M$ emerge as the primary physical processes.
	
	The final step of this section consists in describing the driven quantum evolution in the presence of decoherence channels. To achieve this goal, We use the standard master equation within the Born-Markov approximation, which in Lindblad form for the composite optomechanical density operator takes the form
	
	\begin{align}
	\frac{d\hat{\rho}}{dt}={}& -i\left[ \hat{H},\hat{\rho}\right] +\frac{\kappa}{2}\mathcal{D}\left[ \hat{a} \right] \hat{\rho}  \notag \\
	& +\frac{\gamma }{2}(1+\overline{n}_{m})\mathcal{D}\left[ \hat{b}\right] \hat{\rho}+\frac{\gamma }{2} \overline{n}_{m} \mathcal{D}\left[ \hat{b}^{\dagger }\right] \hat{\rho},  \label{master}
	\end{align}
	with the Lindbladian superoperator term denoted by $\mathcal{D}\left[ \hat{O}\right] = 2\hat{O}\hat{\rho}\hat{O}^{\dagger }-\hat{\rho}\hat{O}^{\dagger }\hat{O}-\hat{O}^{\dagger }\hat{O}\hat{\rho}.$ In the above, notice that we have neglected the reservoir photons number on average $\overline{n}_c$, as the difference of frequency between the light and mechanical spectra make $\overline{n}_c \ll \overline{n}_m$ \cite{Optomechrev} for a finite common environment temperature. Thus, we take into account the dissipative mechanisms of a thermal reservoir with average occupation number $\overline{n}_{m}$ in the mechanical degree of freedom and photon (phonon) decay rate $\kappa (\gamma)$. To obtain the phononic steady-state, we proceed to derive a master equation of the reduced displaced mechanical density operator. To accomplish this, we recast the transformations carried out previously, namely; switching to the interaction picture, performing a proper optomechanical displacement, i.e., $\hat{U}^\dagger \hat{D}^\dagger(\hat{\xi})\cdots \hat{D}(\hat{\xi})\hat{U}$ and, finally, tracing out the optical degrees of freedom. In particular, in the bad-cavity regime, i.e., $\langle \hat{\chi}^{(1)}(\eta) \rangle \ll \kappa$ (the expectation value stands for a specific $M$, such as the $L_M^{(1)}(\eta^2) = 0$), the master equation in the displaced-interaction picture reads as \cite{Effective}
	
	\begin{equation}
	\frac{d\hat{\rho}_{m}}{dt} = \frac{\kappa_\mathrm{eff}}{2}\mathcal{D}\left[\hat{B}^\dagger\right]\hat{\rho}_m + \frac{\gamma}{2}(1+\overline{n}_{m})\mathcal{D}\left[\hat{b}\right]\hat{\rho}_{m} + \frac{\gamma}{2}\overline{n}_{m}\mathcal{D}\left[\hat{b}^{\dagger}\right]\hat{\rho}_{m},  \label{eff}
	\end{equation}
	with the effective damping rate $\kappa_\mathrm{eff} = 4 \langle \hat{\chi}^{(1)}(\eta)\rangle/\kappa.$ In contrast to the full master equation shown in Eq. (\ref{master}), from analyzing the above effective master equation, we readily find the steady state solution under the condition $\kappa_\mathrm{eff} \gg \gamma$, as any initial state $\hat{\rho}_{m} = \sum_{l,m=0}^{M} p_{lm}$ $\left\vert l\right\rangle \left\langle m\right\vert $ is asymptotically driven to $\hat{\rho} (\infty)_{m}\approx \left\vert M\right\rangle \left\langle M\right\vert$. In addition to this, if we consider the effects of a thermal reservoir (originally neglected in our former derivation) with occupation number $\overline{n}_{c}$, the final steady state would be slightly modified as a displaced Fock state given by $\hat{\rho}(\infty )_{m} = \hat{D}(\eta \overline{n}_{c})\left \vert M\right\rangle \left\langle M\right\vert \hat{D}^{\dagger}(\eta \overline{n}_{c})$. 
	
	We confirm our protocol in both the bad and the good cavity regime by solving numerically \cite{QuTIP} the full master equation (\ref{master}), and we investigate the effects of mechanical damping rates and temperature. It is relevant to point out, however, that in the strong optomechanical coupling regime, a more suitable representation for the dissipative dynamics follows a dressed-state master equation \cite{dmaster}. Nonetheless, due to the regime of parameters considered throughout this work, both (standard and dressed-state) master equations lead to the same steady state found in Eq.(\ref{master}). To show that indeed the system approaches the desired Fock state we present a set of complementary measures that confirms the reasoning following from the effective master equation, namely; we compute the fidelity $\mathcal{F}(t) = \sqrt{ \langle M | \hat{\rho}_{m}(\infty) | M \rangle}$ and the associated purity $\mathcal{P}(t) = \mathrm{Tr} \left[ \hat{\rho}_{m}^{2}(\infty)\right]$ \cite{Livro} and, finally, to characterize the nonclassical nature of the state we use the quantity $\mathcal{I}$ \cite{I} defined as
	
	\begin{equation}
	\mathcal{I} = - \frac{\pi}{2} \int \mathrm{d}p \mathrm{d}q W(q,p) \left( \frac{\partial^2}{\partial q^2} + \frac{\partial^2}{\partial p^2} + 1 \right).
	\end{equation}
	
	Here, $W(q,p)$ is the Wigner function and $\mathcal{I}$ goes from 0 (for classical states, like Gaussian and thermal states) to $\mathrm{Tr}\left[\hat{b}^{\dagger }\hat{b}{\rho} (\infty)_{m}\right] = \langle n \rangle$ (average number of excitations in the system) for pure quantum states such as superposition of coherent states, NOON states and Fock states. It is important to stress that, $\mathcal{I}$ is invariant under unitary transformations. Thus, it can readily be computed in the displaced interaction picture chosen by us \cite{I}.
	
	\section{Numerical results}
	
	\begin{figure}
		\centering \includegraphics[width=\linewidth]{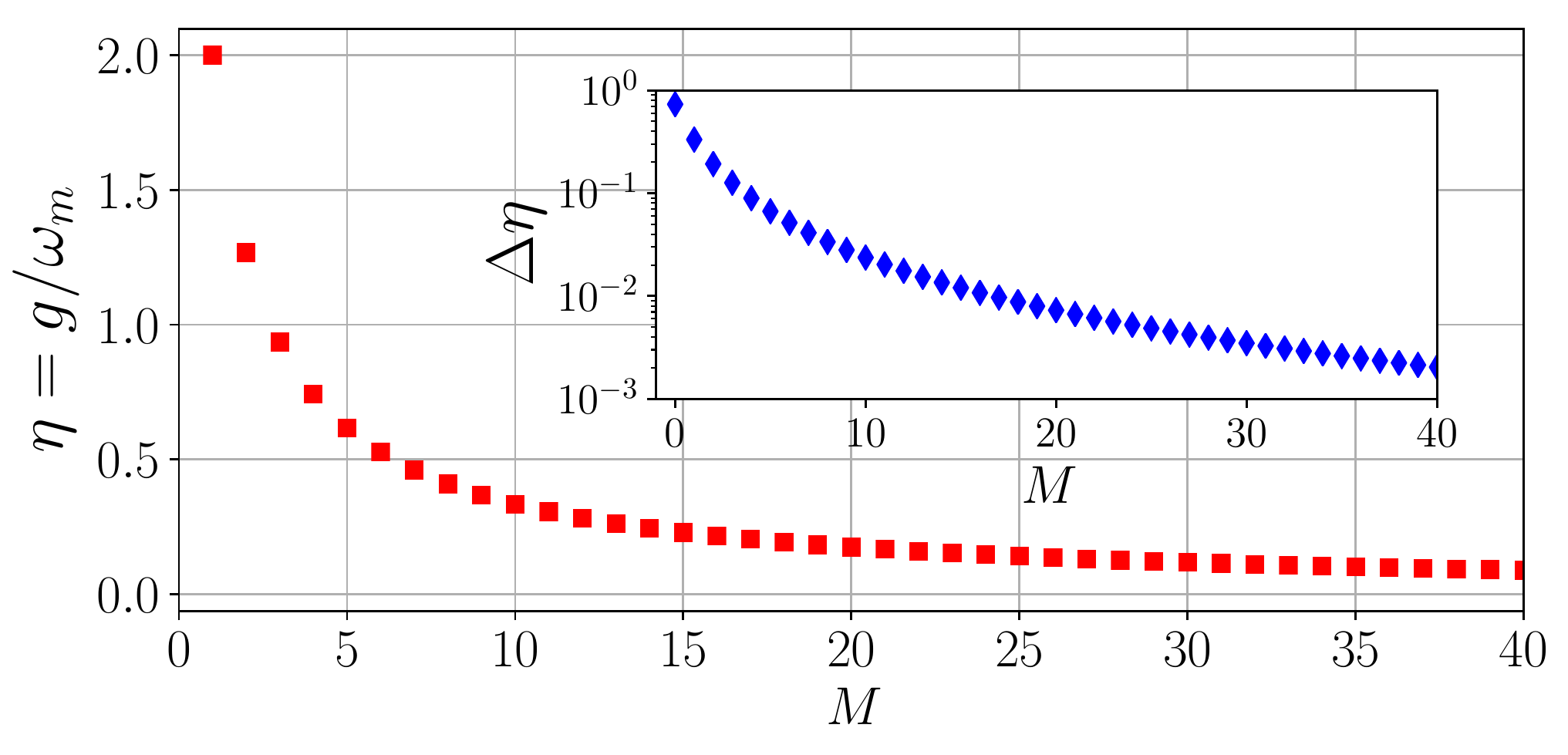}
		\caption{Required optomechanical coupling $\eta$ to make the Laguerre polynomial vanish [i.e., $L_M^{(1)}(\eta^2) = 0$] as a function of Fock number states $|M \rangle$. The inset plot shows the difference between neighboring coupling values in order to prepare adjacent Fock states, i.e., $\Delta \eta = |\eta_{M+1} - \eta_M|$, showing how resolvable $\eta$ must be to prepare each state.}
		\label{M-Laguerre}
	\end{figure}
	
	Let us begin our analysis by targeting some specific Fock states, for which we have chosen $|M = 5 \rangle$ and $|M = 10 \rangle$ only for illustrative purposes. Certainly, other Fock states can also be prepared, as long as the zeroes of the generalized Laguerre polynomial can be resolved and achieved in our optomechanical setup. To show the connection between the target state $|M\rangle$ and the necessary $\eta$ to achieve $|M\rangle$, we have depicted in Fig. \ref{M-Laguerre} the $\eta$ value for which the Laguerre polynomial vanishes (its first zero), i.e., $L_M^{(1)}(\eta) = 0$. Notice that production of "small" Fock number states for the mechanical degree of freedom might represent a challenging parameter region. For example, for $M \leq 2$ the single-photon coupling is comparable to the mechanical frequency $g \geq \omega_m$. However, several setups have been recently proposed to reach strong photon-phonon nonlinearities, and therefore paving the way to realize "small" and considerable "large" phononic Fock number state experimentally. For instance, $\eta \sim 0.2$ (moderately strong, which in turn would enable the production of states with $M \sim 15$) for the optomechanical interaction has already been exceeded in a nanostring optomechanical cavity \cite{Krause2015} and also in a novel sliced photonic crystal nanobeam scheme with $\eta > 1$ \cite{Leijssen2015}. Other systems involve; membrane-in-the-middle architectures \cite{Neumeier2018}, and high-finesse optomechanical microcavities \cite{Vanner2011}, to name a few. Modest improvements in some on-chip systems can be carried on \cite{Safavi2011, ChanAlegre2011} e.g., by decreasing both $\omega_m$, as $g \propto \omega_m^{-3/2}$, and $\kappa$ by just 1-2 orders of magnitude. Promising recent proposals to exceed the required $\eta$ have been made \cite{Xuereb2012, Kaviani2015}, and also in levitated helium drop systems \cite{helium}. In the inset plot of Fig. \ref{M-Laguerre}, we show the significance of the resolution in $\eta = g/\omega_m$; as $M$ increases, the difference between the required $\eta$ to prepare a specific steady-state $M$ decreases substantially. Hence, a slight deviation from $\eta_M$ might turn into generating undesired phonon states.
	
	\begin{figure}[t]
		\centering \includegraphics[width=0.9\linewidth]{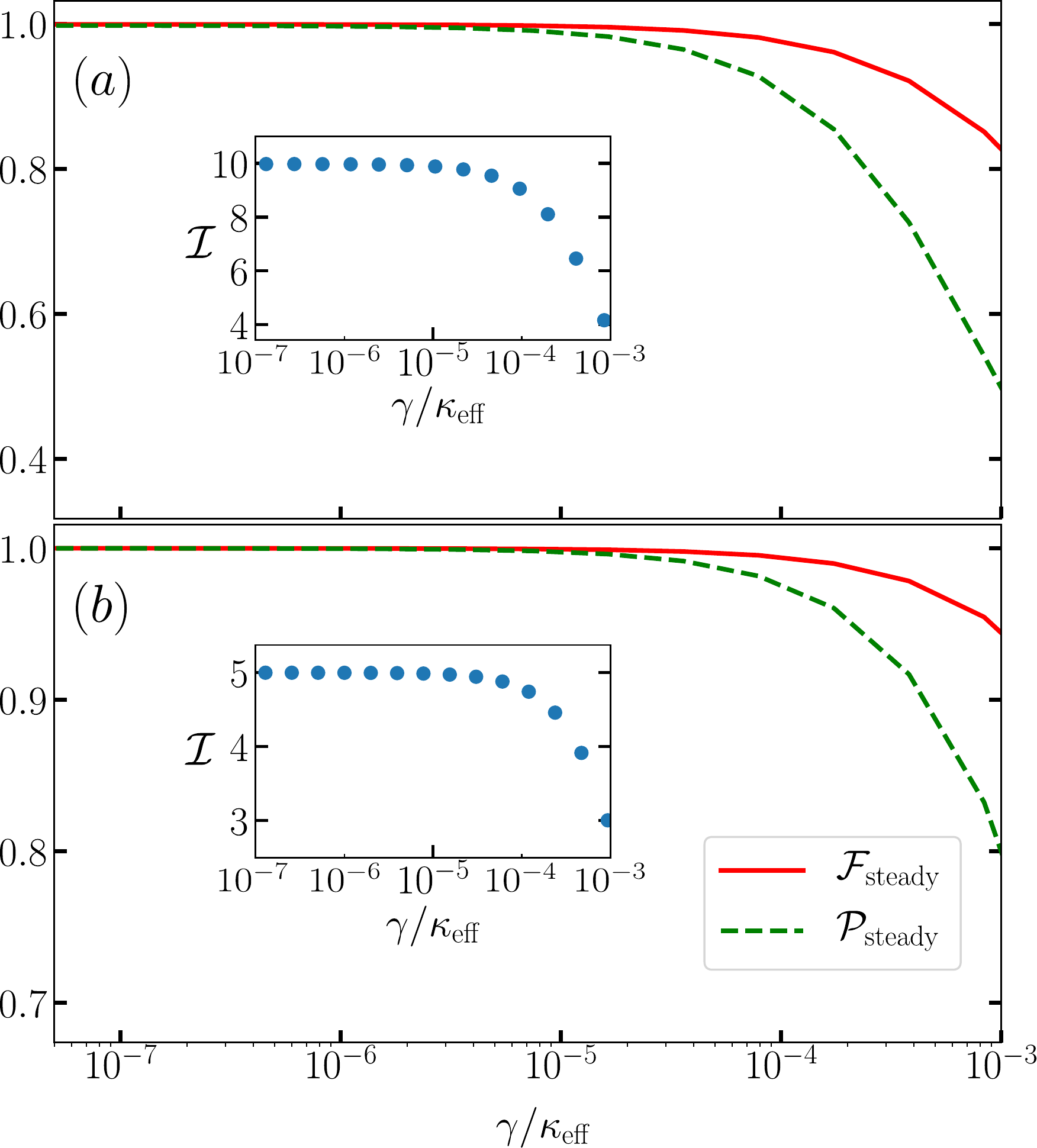}
		\caption{Fidelity and purity for two different phonon target states as a function of the ratio $\gamma/\kappa_{\mathrm{eff}}$. In the top panel (a), we consider $|M=10\rangle$, whereas (b) shows the target state $|M=5\rangle$. The inset plot shows the nonclassicality quantity $\mathcal{I}$ in the same $\gamma/\kappa_{\mathrm{eff}}$ interval; $\eta$ has been calculated to supress the Laguerre polynomial, and we fixed $\overline{n}_m = 0.3.$}
		\label{fig:fig3}
	\end{figure}
		\begin{figure}[t]
		\centering \includegraphics[width=0.9\linewidth]{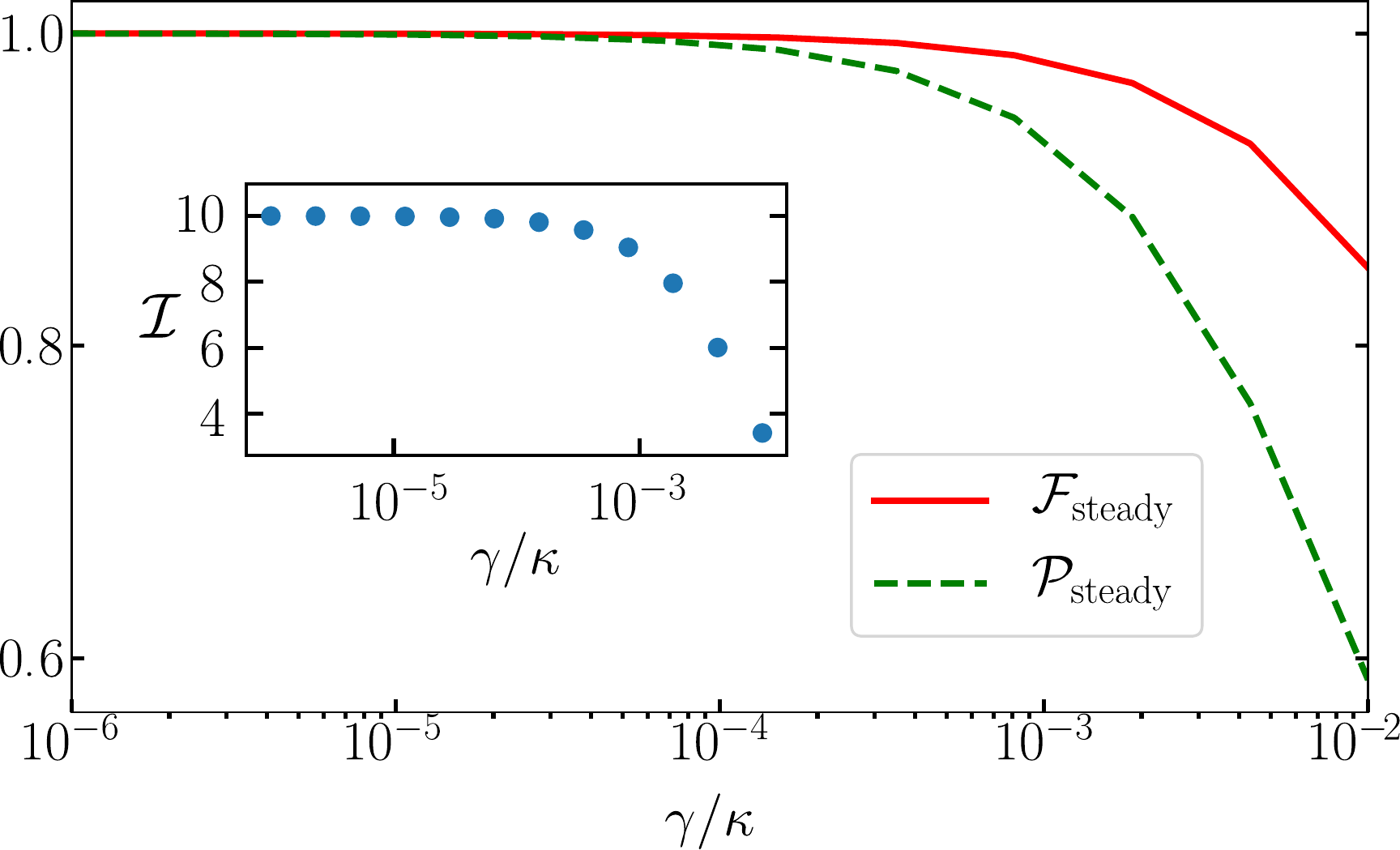}
		\caption{Fidelity and purity for Fock phonon state production $|M = 10\rangle$ in the good-cavity limit $\langle \hat{\chi}^{(1)}(\eta) \rangle = \kappa$. Again, the inset shows the non-classicality $\mathcal{I}$ in the same interval of $\gamma/\kappa$.}
		\label{fig:fig4}
	\end{figure}
	
	\begin{figure*}[t]
		\includegraphics[width=\linewidth]{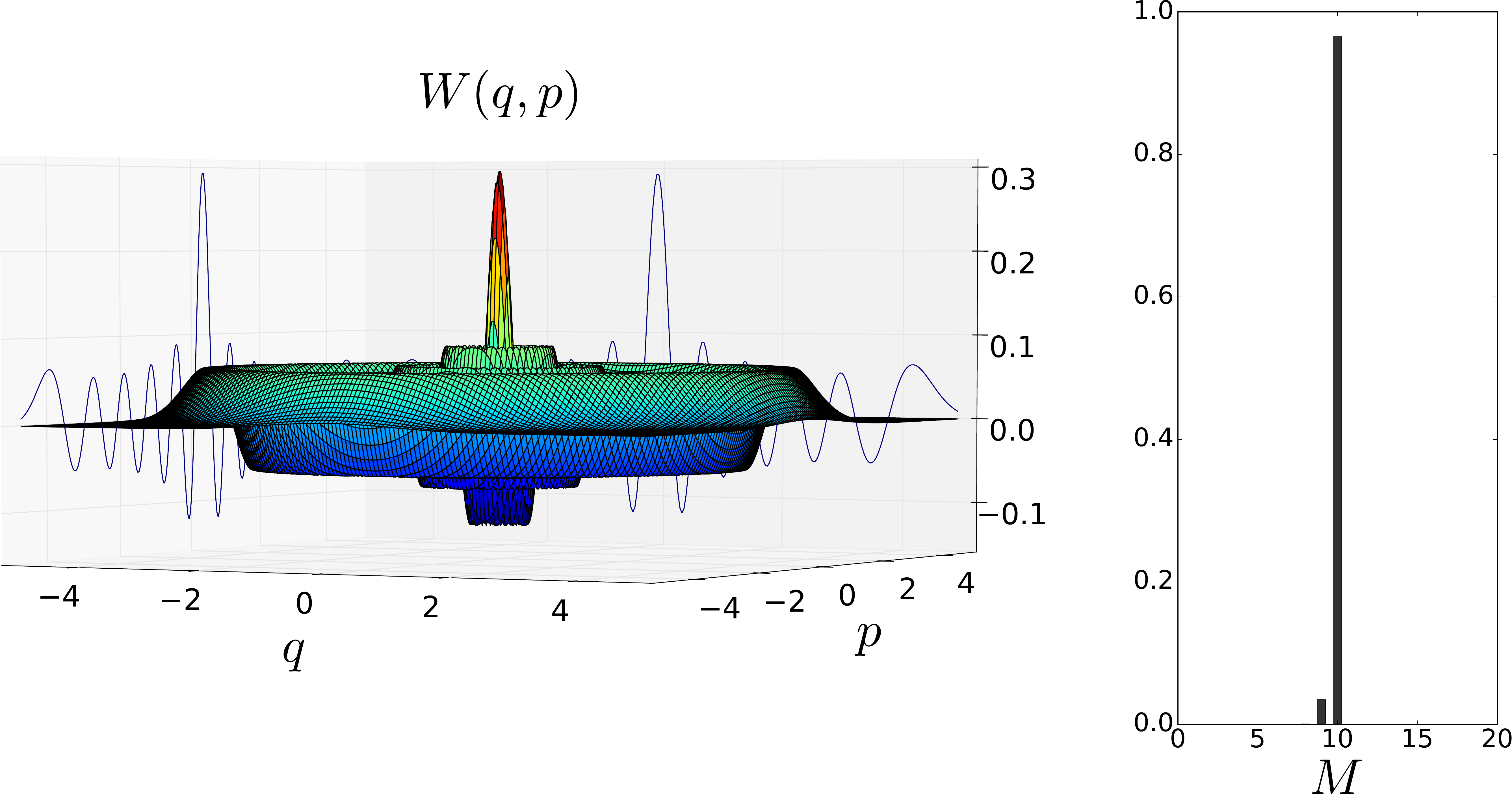}
		\caption{In the left panel we depict the Wigner quasi-probability function for the dissipative engineered state $|M = 10\rangle$ in the good-cavity regime $\langle \hat{\chi}^{(1)}(\eta) \rangle = \kappa$ to $\gamma/\kappa=10^{-3}$. In the right panel, we show the phonon number occupancy, where the most probable state is centered at $M = 10$, depicting its high fidelity with the target phonon state $|10 \rangle$.}
		\label{fig:fig5}
	\end{figure*}
	In what follows, we investigate up to which values in the dissipative channels $\{\gamma, \kappa, \kappa_\mathrm{eff}\}$ our Fock state production scheme can be accommodated. For the sake of simplicity, we will neglect $\overline{n}_c$ in our simulations, whereas the mean phonons occupancy number on average will be fixed as $\overline{n}_m = 0.3$ ---a value for thermal phonons as low as $\overline{n}_m \approx 0.3$ can be achieved experimentally in mechanical resonators operating in the microwave regime at milliKelvin temperatures \cite{OConnell2010}. First, we present our results for the bad-cavity regime [$\langle \hat{\chi}^{(1)}(\eta) \rangle \ll \kappa$] in Fig. \ref{fig:fig3}. There, we have depicted the fidelity and purity for the dissipative production of $|5\rangle$ (bottom panel) and $|10\rangle$ (top panel) Fock states as a function of the ratio $\gamma/\kappa_\mathrm{eff}$, where $\kappa_\mathrm{eff} = 4 \langle \hat{\chi}^{(1)}(\eta)\rangle/\kappa.$ As observed in the main plot, the generated Fock steady-state can be accommodated up to values of $\sim 10^{-4}$, where the $\mathcal{P}_\mathrm{steady} \approx 0.9$ and $\mathcal{F}_\mathrm{steady} \approx 0.96$. In their respective insets plot of Fig. \ref{fig:fig3}, we show the nonclassicality quantity $\mathcal{I}$. For values $\gamma/\kappa_\mathrm{eff} < 10^{-4}$ it can be seen that $\mathcal{I} \approx M$, hence, reinforcing the fact that the preparation of nonclassical states for the mechanical degree of freedom is feasible.

	Finally, we would like to show the validity of our dissipative scheme under the good cavity operational regime in the single-photon strong coupling regime [$\langle \hat{\chi}^{(1)}(\eta) \rangle \sim \kappa$]. To exhibit these results, we proceed to solve the full master equation shown in Eq. (\ref{master}).In Fig. \ref{fig:fig4}, we observed that the production of the state $|M = 10\rangle$, in contrast to the bad cavity regime, can be reached with one order of magnitude higher within the dissipative ratio $\gamma/\kappa$. In other words, for similar wanted fidelities ($> 0.9$) and purities ($> 0.95$), the generation of nonclassical phonon states in the good cavity regime stands as more robust against decoherence, in contrast to the bad cavity limit. Nonetheless, in both cases, $\gamma/\kappa_\mathrm{eff} \sim 10^{-4}$ and $\gamma/\kappa \sim 10^{-3}$, can nowadays be attained experimentally \cite{Optomechrev}. In Fig. \ref{fig:fig5}, to exhibit the quantumness of the generated Fock state, we present the Wigner quasi-probability distribution for $|M = 10\rangle$ in the good cavity regime, together with the phonon number occupancy.

	\section{Final remarks}
	We present an on-demand dissipative scheme to prepare phononic Fock number states in the nonlinear optomechanical single-photon strong interaction. Specifically, we have studied a system composed of a standard laser-driven cavity, where no-linearization of the Hamiltonian has been performed. Moreover, when we reach the effective Hamiltonian, two main physical processes arise. On the one hand, although an external laser dynamically drives the cavity, only 0 and 1 intracavity photons transitions take place, in the same manner as in the photon blockade effect. On the other hand, a precise selection of the optomechanical coupling strength makes the associated Laguerre polynomial vanish for a specific $M$ phonon number state, i.e., a dark state for the mechanical degree of freedom. This last process can be viewed, hence, as slicing the Hilbert space of the phonon degree of freedom via dissipative engineering. We readily notice that our proposal requires strong optomechanical interactions $g > \omega_m$ for $M \leq 5$, whereas for larger phonon productions works for strong-moderate optomechanical couplings $g/\omega_m \sim 0.4$. We have justified requiring the strong-moderate operational regime with novel optomechanical setups, where the single-photon coupling has been achieved or exceeded. For instance, the single-photon strong optomechanical coupling has been attained experimentally in a BEC-cavity system \cite{BEC}. We show that our results are promising both in the good and the bad cavity regimes, with fidelities exceeding $\mathcal{F} > 0.9$ and purities above $\mathcal{P} > 0.95$. The 'quantumness' of the Fock phonon steady-state has been provided with a numerical non-demanding nonclassicality measurement ($\mathcal{I}$), and also with the Wigner quasi-probability distribution. Finally, in addition to the preparation of Fock states, other applications may arise from the present protocol, such as the production of entangled steady state in optomechanics arrays.

	\section{Funding Information}
	G. D. de M. N. and V. M. acknowledge funding from the Chinese Postdoctoral Science Fund 2018M643436 and 2018M643435, respectively. V. F. T. acknowledges financial support from CAPES. This work was supported by CAPES, CNPq (Grant 206224/2014--1 (PDE)) and FAPEMIG.
	
	\section*{Acknowledgments}
	
	V. F. T. would like to acknowledge C. Cherubim, R. F. Rossetti, H. S. de Ara\'{u}jo, M. H. Y. Moussa and O. S. Duarte for valuable discussions. 
	
	\section*{Author Contributions}
	
	G. D. de M. N. and V. F. T designed the research; G. D. de M. N., V. M., V. F. T. performed the numerical simulations. All authors provided suggestions, discussed the content, reviewed and edited the manuscript.

	
	
	
\end{document}